\documentclass[graybox]{svmult}

\usepackage{mathptmx}       
\usepackage{helvet}         
\usepackage{courier}        
\usepackage{type1cm}        
\usepackage{makeidx}         
\usepackage{graphicx}        
\usepackage{multicol}        
\usepackage[bottom]{footmisc}

\newcommand{\apj}{ApJ}
\newcommand{\apjl}{ApJL}
\newcommand{\mnras}{MNRAS}
\newcommand{\aap}{A\&A}

\makeindex             

\begin{document}

\title*{MHD models of Pulsar Wind Nebulae}
\author{Niccol\`o Bucciantini}
\institute{N.~Bucciantini \at NORDITA, Roslagstullsbacken 23, 106 91 Stockholm,
  Sweden, \email{niccolo@nordita.org}}
%
%
\maketitle

\abstract*{Pulsar Wind Nebulae  (PWNe) are bubbles or relativistic plasma that form when the
  pulsar wind is confined by the SNR or the ISM. Recent observations
  have shown a richness of emission features that has driven a renewed
interest in the theoretical modeling of these objects. In recent years
a MHD paradigm has been developed, capable of reproducing almost all of
the observed properties of PWNe, shedding new light on many old
issues. Given that PWNe are perhaps the nearest systems where processes
related to
relativistic dynamics can be investigated with high accuracy, a
reliable model of their behavior is paramount for a correct
understanding of high energy astrophysics in general. I will review the present
status of MHD models:
what are the key ingredients, their successes, and open
questions that still need further investigation.}

\abstract{Pulsar Wind Nebulae  (PWNe) are bubbles or relativistic plasma that form when the
  pulsar wind is confined by the SNR or the ISM. Recent observations
  have shown a richness of emission features that has driven a renewed
interest in the theoretical modeling of these objects. In recent years
a MHD paradigm has been developed, capable of reproducing almost all of
the observed properties of PWNe, shedding new light on many old
issues. Given that PWNe are perhaps the nearest systems where processes
related to
relativistic dynamics can be investigated with high accuracy, a
reliable model of their behavior is paramount for a correct
understanding of high energy astrophysics in general. I will review the present
status of MHD models:
what are the key ingredients, their successes, and open
questions that still need further investigation.}

\section{Introduction}
\label{sec:1}
When the ultra-relativistic wind from a pulsar interacts with the
ambient medium, either the SNR or the ISM, a bubble of non-thermal
relativistic particles and magnetic field, known as Pulsar Wind
Nebula or ``Plerion''  (PWN), is formed. The Crab Nebula is
undoubtedly the best example of a PWN, and it is often considered the
prototype of this entire class of objects, to the point that models of
PWNe are, to a large extent, based on what is known in this single
case. The first theoretical model of the structure and the dynamical
properties of PWNe was presented by Rees \& Gun \cite{ree74},
further developed in more details by Kennel \& Coroniti
\cite{ken84a,ken84b} (KC84 hereafter), and is based on a relativistic
MHD description. 

The MHD paradigm is based on three key assumptions: 
\begin{itemize}
\item that the larmor radii of the particles is much smaller than the
  typical size of the nebula, and particles are simply advected with
  the magnetic field. This is true up to energies of order of the
  pulsar's voltage, where the larmor radius becomes comparable with the typical
  size of the system.
\item That radiative losses are negligible, or at least that
  they can be accounted for by renormalizing the pulsar spin-down
  luminosity. This again can be proved to be true in the case of Crab
  Nebula (and to some extent also in other systems with good spectral
  coverage), where the synchrotron spectrum shows that the particles carrying
  the bulk of the energy have a typical lifetime for synchrotron
  cooling longer than
  the age of the nebula.
\item That we are dealing with almost pure pair plasma, and dispersive
  or hybrid effects (separation of scales) due to the presence of heavier ions are
  absent. While there is no direct evidence for the absence of ions,
  standard pulsar wind theory, and the success of the MHD model of
  PWNe suggest that, from a purely dynamical point of view, there is
  no need for this extra component.
\end{itemize} 

\begin{figure}
\label{fig:4}
\includegraphics[scale=0.2]{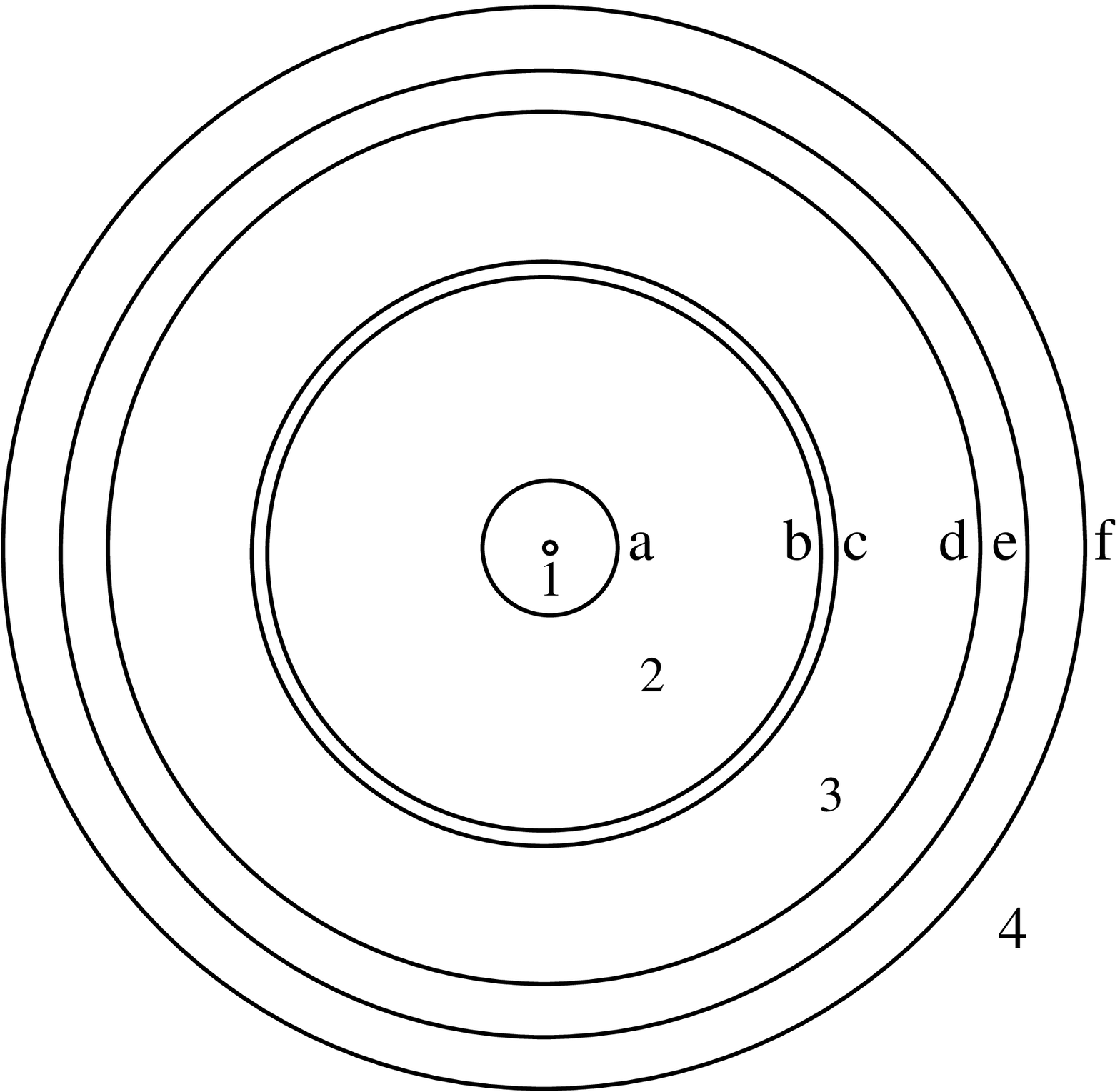}\hspace{0.5truecm}\includegraphics[scale=0.35]{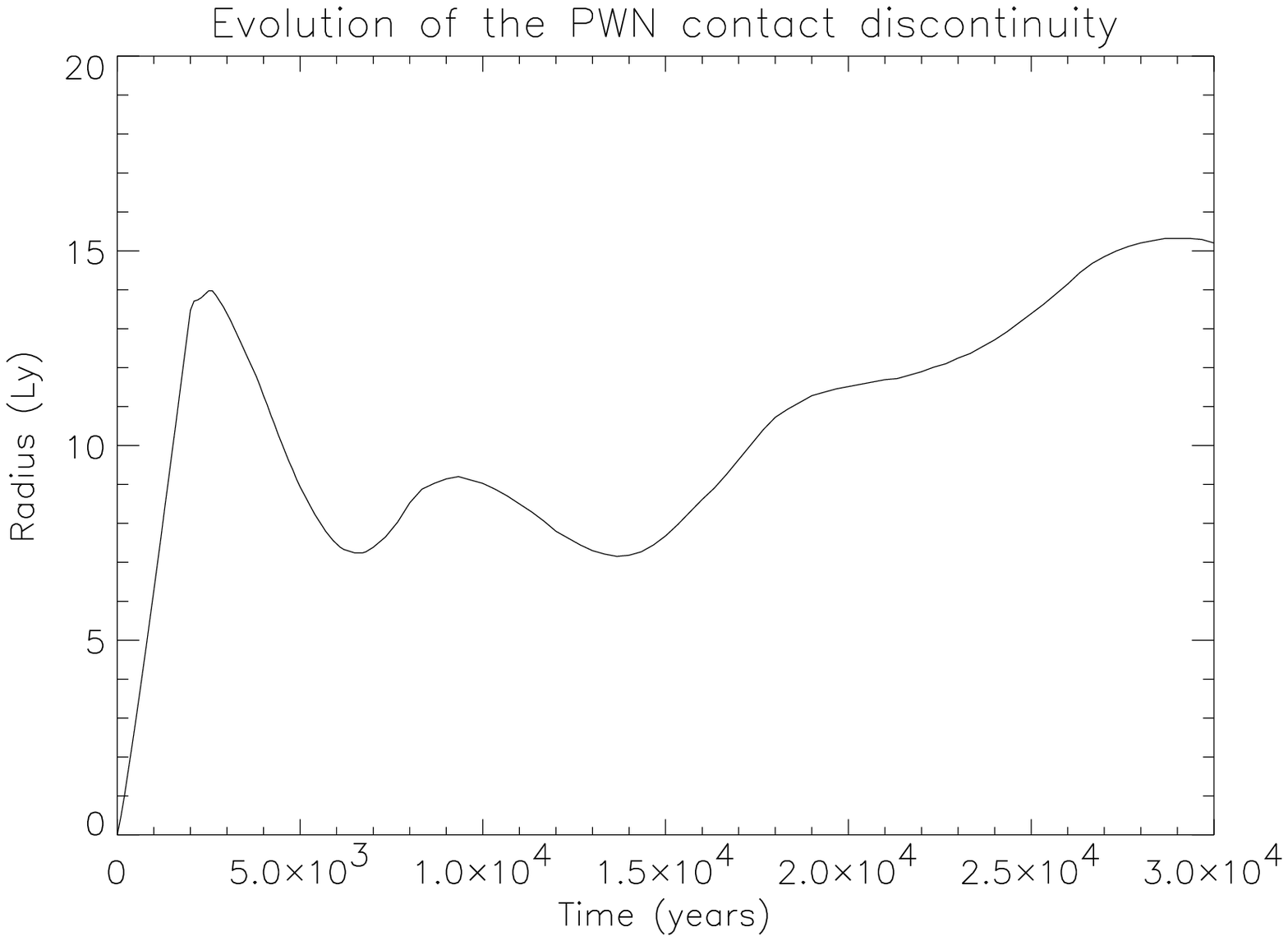}
\caption{Left picture: schematic representation of the  global
  structure of the PWN in the first phase of its evolution inside a
  SNR.  From the center the various regions are: 1- the  relativistic
  pulsar wind, 2- the hot magnetized bubble responsible for the non
  thermal emission, 3- the free expanding ejecta of the SNR, 4- the
  ISM. These regions are separated by discontinuities: a- the wind
  termination shock, b- the contact discontinuity between the hot
  shocked pulsar material and the swept-up SNR ejecta, c-the front
  shock of the thin shell expanding into the ejecta, d- the reverse
  shock of the SNR, e- the contact discontinuity separating the ejecta
  material from the compressed ISM, f- the forward SNR shock. Right
  picture: evolution of the PWN size, from free-expansion to sedov
  phase (from \cite{me03}).}
\end{figure}

In it simplest form \cite{ree74} the MHD model of PWNe can be
summarized as follow (see Fig.~\ref{fig:4}): the ultra-relativistic pulsar wind is confined
inside the slowly expanding SNR, and slowed down to non relativistic
speeds in a strong termination shock (TS). At the shock the plasma is
heated, the toroidal magnetic field of the wind is compressed, and
particles are accelerated to high energies. These high energy particles
and magnetic field produce a post-shock flow which expands at a non
relativistic speed toward the edge of the nebula.

Despite its simplicity the MHD model can explain many of the observed
properties of PWNe, and until now no observation has been presented
that could rule it out.  The presence of an under-luminous region,
centered on the location of the pulsar, is interpreted as due to the
ultra-relativistic unshocked wind. Polarization measures
\cite{wil72,vel85,sch79,hic90,mic91,dod03,kot06,hes08} show that emission is highly
polarized and the nebular magnetic field is mostly toroidal, as one
would expect from the compression of the pulsar wind, and it is
consistent with the inferred symmetry axis of the system. The pressure
anisotropy associated to the compressed nebular toroidal magnetic field
\cite{beg92,van03}, explains the elongated axisymmetric shape of many
PWNe ({\it i.e.} Crab Nebula, 3C58). The MHD flow from the TS to the
edge of the nebula also leads to the prediction that PWNe should appear
bigger at smaller frequencies: high energy X-rays emitting particles
are  present only in the vicinity of the TS, having a shorter
lifetime for synchrotron losses, compared to Radio particles which
fill the entire volume, having negligible losses on the age of the
nebula. This increase in size at smaller frequencies is observed in
the Crab Nebula \cite{ver93,bie97,ban98}. 

However one must bear to mind that not all properties of PWNe can be
explained within the MHD framework, which, ultimately, only provides a
description of the flow dynamics. For example, the acceleration of
particles at the TS that accounts for the continuous, non-thermal,
very broad-band spectrum, extending from Radio to X-rays
\cite{ver93,ban99,wei00,wil01,mor04}, is usually assumed as given. The
MHD model provides no hint to the reason why the injection spectrum
looks like a broken powelaw, with no sign of a Maxwellian component at
lower energies. Moreover the MHD description might prove faulty if
applied to particles responsible for the emission in the 10-100 MeV
band, whose larmor radii are comparable to the size of the TS,
and can lead to wrong conclusion on their expected behavior.

The MHD model of PWNe has been used, by comparing observations with
the predictions of numerical simulations, to  constrain some of the
properties of the pulsar wind, at least at the distance of the
TS. While it is not possible to derive the Lorentz factor of the wind,
or its multiplicity,
it is possible to constrain the ratio between Poynting flux and
kinetic energy, the latitudinal dependence of the energy flux, and the
presence of a dissipated equatorial current sheet. This shows that
nebular properties can be used to derive informations on the
conditions of the pulsar wind at large distances.

\section{Jet-Torus structure and Inner flow properties}
\label{sec:2}

Let start our discussion from young objects. The Crab Nebula, 3C58, MSH
15-52, G21.5 all belongs to this group. Younger objects are the most
well studied and perhaps the ones for which the MHD models have provided
the greatest insight. These systems are characterized by a simple
interaction with the confining SNR, they are bright, we have broad
band data, and the pulsar proper motion can be
neglected. Older systems are often subject to a much more complex
interaction with the SNR, they are affected by the pulsar proper
motion, and usually lack the deep observational coverage of the younger
counterparts. For this reasons, models of old objects have also
progressed far less than for young ones, and the agreement with
observations is mostly qualitative. We will leave a description of MHD
models of the evolution of PWNe to Sec.~\ref{sec:3}.

The KC84 model has been for a long time the reference for the
understanding of young PWNe, with only minor theoretical
developments. Things have changed recently thanks to high
resolution optical and X-rays images from HST, CHANDRA and XMM-Newton,
that have have shown that the properties of the emission at high energy cannot be
explained within a simplified one dimensional model. This 
refers not just  to the geometrical features that are observed, but in
practice to all aspect of X-ray emission.  

\begin{figure}
\label{fig:3}
\includegraphics[scale=.25]{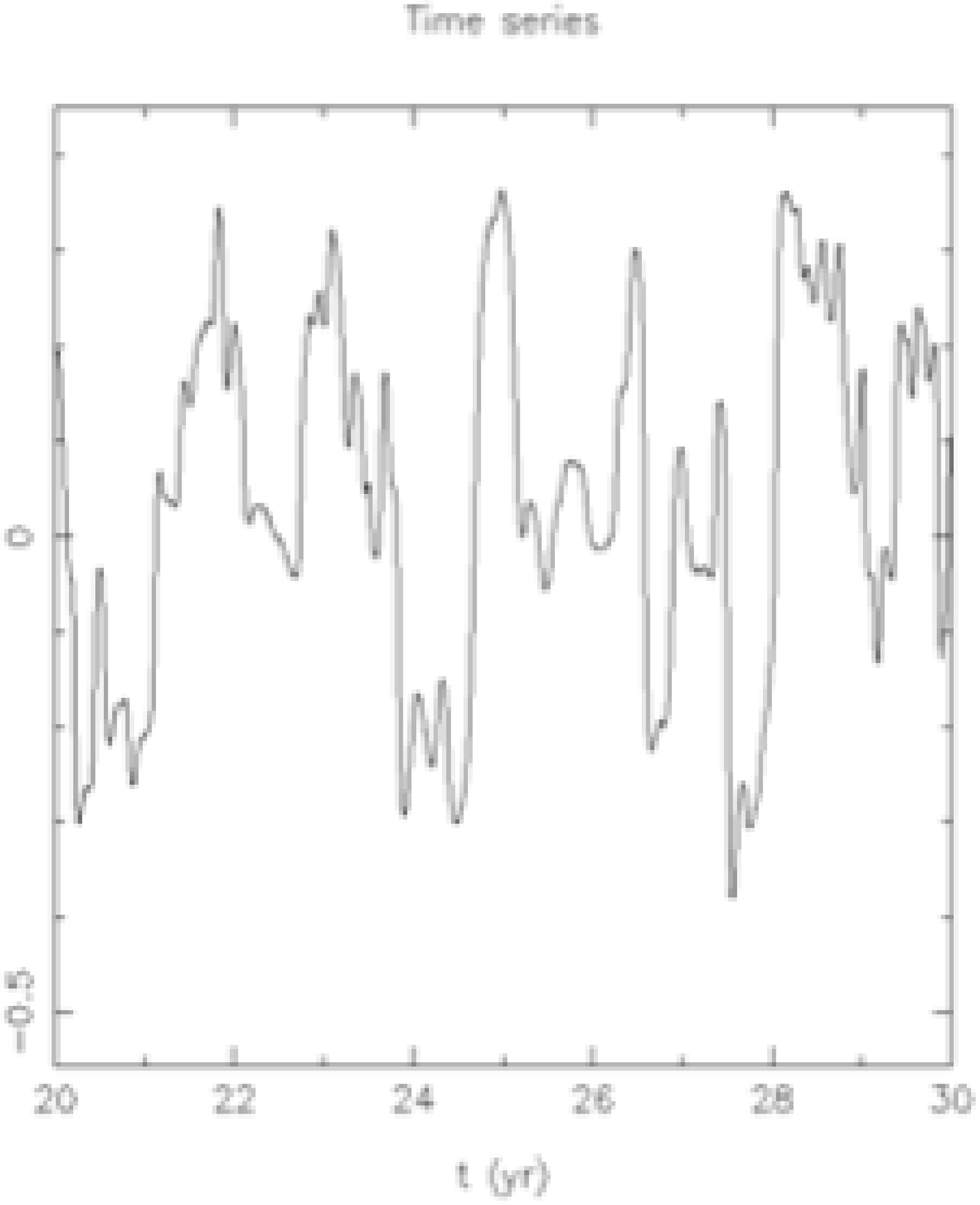}\includegraphics[scale=.3]{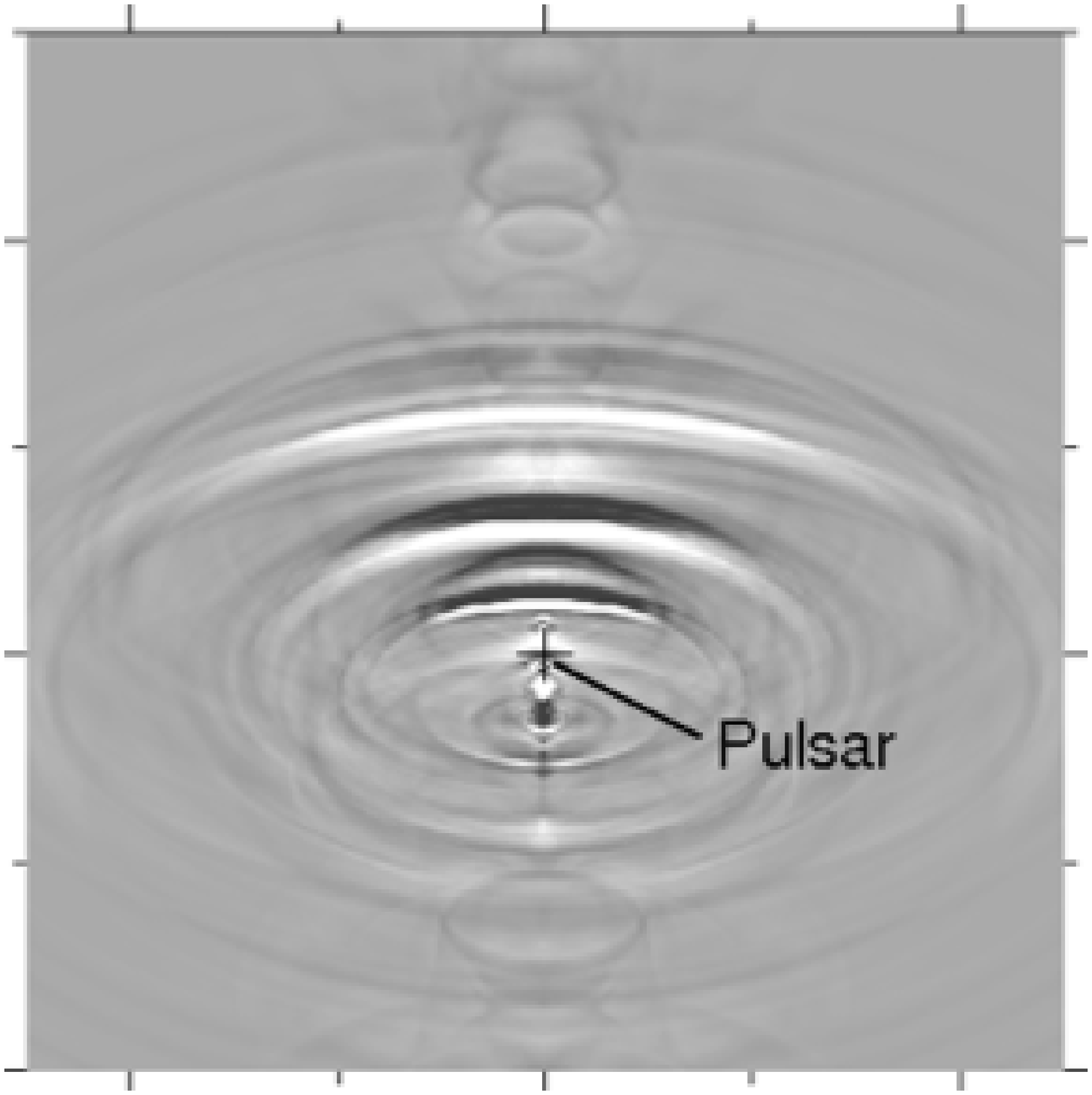}
\caption{Variability in the wisps region in MHD \cite{cam09}. Left
  figure: variability at a selected point on axis, note the typical
  2-years long cycle and also the shorter timescale variations. Right
  figure:
  variability of the nebula obtained by subtracting two images at
  2-year distance. The outgoing wave pattern is easily seen. Compare
  with optical images from \cite{hes02}.}
\end{figure}

These new data
show that the inner region of young PWNe is characterized by a complex
axisymmetric structure, generally referred as {\it jet-torus
  structure} (Fig.~\ref{fig:3}). First observed in Crab
\cite{hes95,wei00}, it has subsequently been detected in many other
PWNe
\cite{got00,gae01,hel01,pav03,gae02,lu02,rom03,sla04,cam04,rom05}, to
the point that the common consensus is that, with deep enough
observations, it should always be detected. This structure is characterized by an emission torus, in what is
thought to be the equatorial plane of the pulsar rotation, and,
possibly, a series of multiple arcs or rings, together with a central
knot, almost coincident with the pulsar position, and one or two
opposite jets along the polar axis, which seem to originate close to
the pulsar itself. Even if the existence of a main torus could be
qualitatively explain as a consequence of a higher equatorial energy
injection \cite{bog02a} it is not possible to reproduce quantitatively
the observed luminosity. Shibata et al. \cite{shi03} were the first to
point that, the difference in brightness between the front and back
sides of the torus in Crab Nebula, requires a post-shock flow velocity
$\sim 0.4-0.5 c$, much higher than what expected for subsonic
expanding flows. The same conclusion applies to all the other systems
where a torus is observed. The existence of an inner ring, detached from the
torus, and of the knot which seemed to be located inside the wind
region, are incompatible with the assumption of a smooth flow from the
TS. From a theoretical point of view however, the most interesting
feature is the jet \cite{lyu01}, because theoretical
\cite{beg94,bes98} and numerical \cite{con99,bog01,gru05,kom06,me06}
studies of relativistic winds from pulsars have shown no presence of
collimated energetic outflow. To this, one must add other observed
properties, like the X-rays photon index maps of Crab Nebula, Vela and
Kes 75 \cite{mor04,pav07,ng08}, which hardens moving from the inner ring toward
the main torus, while steepening due to synchrotron losses is
expected, and the relatively large size of the X-ray nebula in Crab
compared to the radio \cite{ama00}. The fact that the symmetry axis of the jet-torus corresponds
to the major axis of the nebula, leads immediately to the conclusion
that the toroidal magnetic field is paramount in shaping the inner flow.

\begin{figure}
\label{fig:1}
\includegraphics[scale=.4,angle =270]{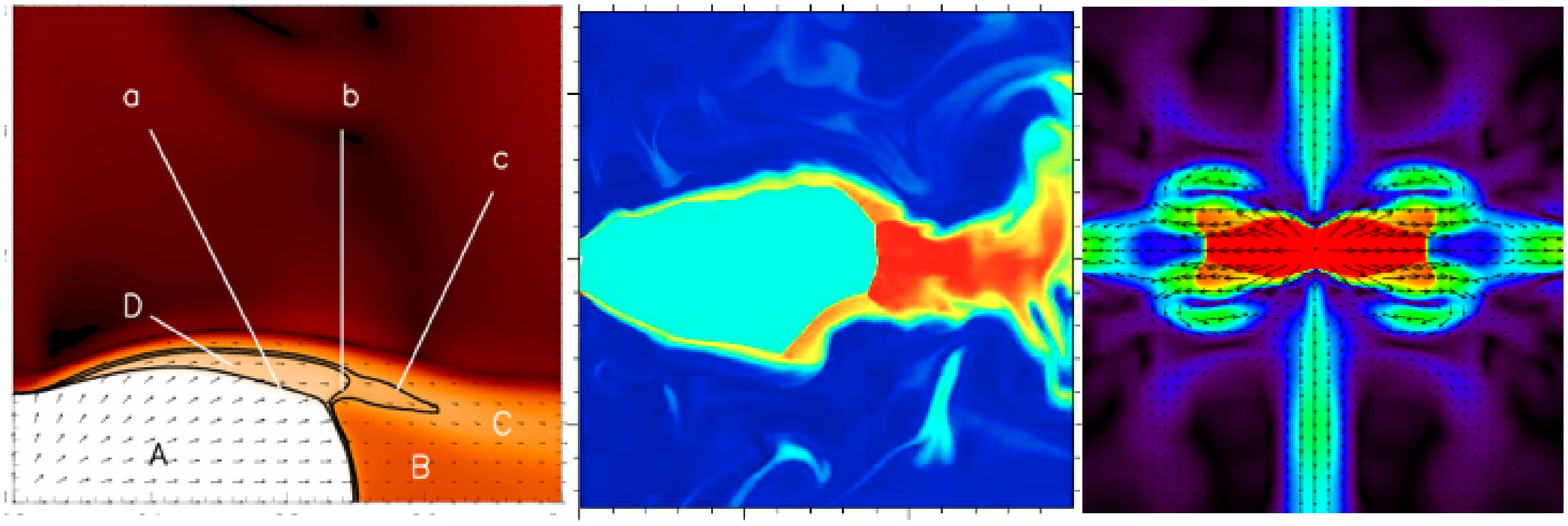}
\caption{From left to right: structure of the post shock flow in PWNe
  \cite{ldz04}, the funneling of the wind (A) into an equatorial flow
  (B/C) is clearly evident. Same structure showing the complex flow
  dynamics that develops downstream of the shock and the corrugation
  of the shock surface which manifests itself as time variability at
  high energy \cite{cam09}. Numerical result of the internal dynamics
  in the body of the PWN \cite{ldz04} where  the flow is 
  diverted back toward the axis by the magnetic hoop stresses, and is
  collimated into a jet.
 }
\end{figure}

The keys in understanding the jet-torus structure are the
magnetization and energy distribution in the pulsar wind. It has been
known for a long time \cite{mic73}, and has been recently confirmed
with numerical simulations \cite{bog01,kom06,me06}, that far from the
Light Cylinder a higher equatorial energy flux is expected. It is this
particular latitudinal distribution of the pulsar spin-down luminosity
which naturally produces an oblate TS with a cusp in the polar region
\cite{bog02a,bog02b}, giving rise to a complex post-shock
dynamics. The obliquity of the TS at higher latitudes, forces the flow
in the nebula toward the equator with speeds $\sim 0.3-0.5
c$. Hoop-stresses are more efficient in the mildly relativistic flow,
and the collimation of a jet occur in the post shock region
\cite{lyu02,kan03}. The evident complexity of this scenario makes clear
that the only possible way to proceed require the use of
 efficient and robust  numerical schemes for relativistic MHD
\cite{kom99,ldz03,gam03}. Thanks to numerical simulations 
this qualitative picture has been developed into a quantitative model
which has been
successfully validated against observations. 

The starting point of the MHD model is the structure of the force-free
pulsar wind: the energy flux in the wind has a strong latitudinal
dependence of the form $L(\theta)=L_o(1+\alpha\sin{(\theta)})$, where
$\alpha$ is a measure of the pole-equator anisotropy, while the
magnetic field in the wind $B(\theta)\propto \sin{(\theta)}$. Various
numerical simulations of the interaction of such wind with the SNR
ejecta have been presented \cite{kom04,ldz04,bog05,ldz06,vol09,cam09}: the result
of the anisotropic energy distribution in the wind is that almost all
of the downstream plasma is deflected toward the equatorial plane, and
flow channels with velocity $\sim 0.5c$ can form, the value expected
in order to justify the luminosity distribution in the torus of the
Crab Nebula \cite{shi03}. Shear and instabilities tend to destroy this
collimated equatorial flow before it reaches the edge of the nebula,
however a bulk equatorial motion survives to distances corresponding
to the location of the torus. It is this flow inside the nebula, with
speed in excess of $c/3$ that advect freshly injected particles to
larger distances, giving rise to a more extended X-ray nebula, that
the simple 1D model would predict\cite{ama00}. The post shock flow is
independent on the specific values of Lorentz
 factor or density distribution, but it is only determined by the
 puslar spin-down energy distribution, and the nebular dynamics cannot be
 used to constrain the value of the wind Lorentz factor or the
 multiplicity in the wind. 

\begin{figure}
\label{fig:2}
\includegraphics[scale=.6]{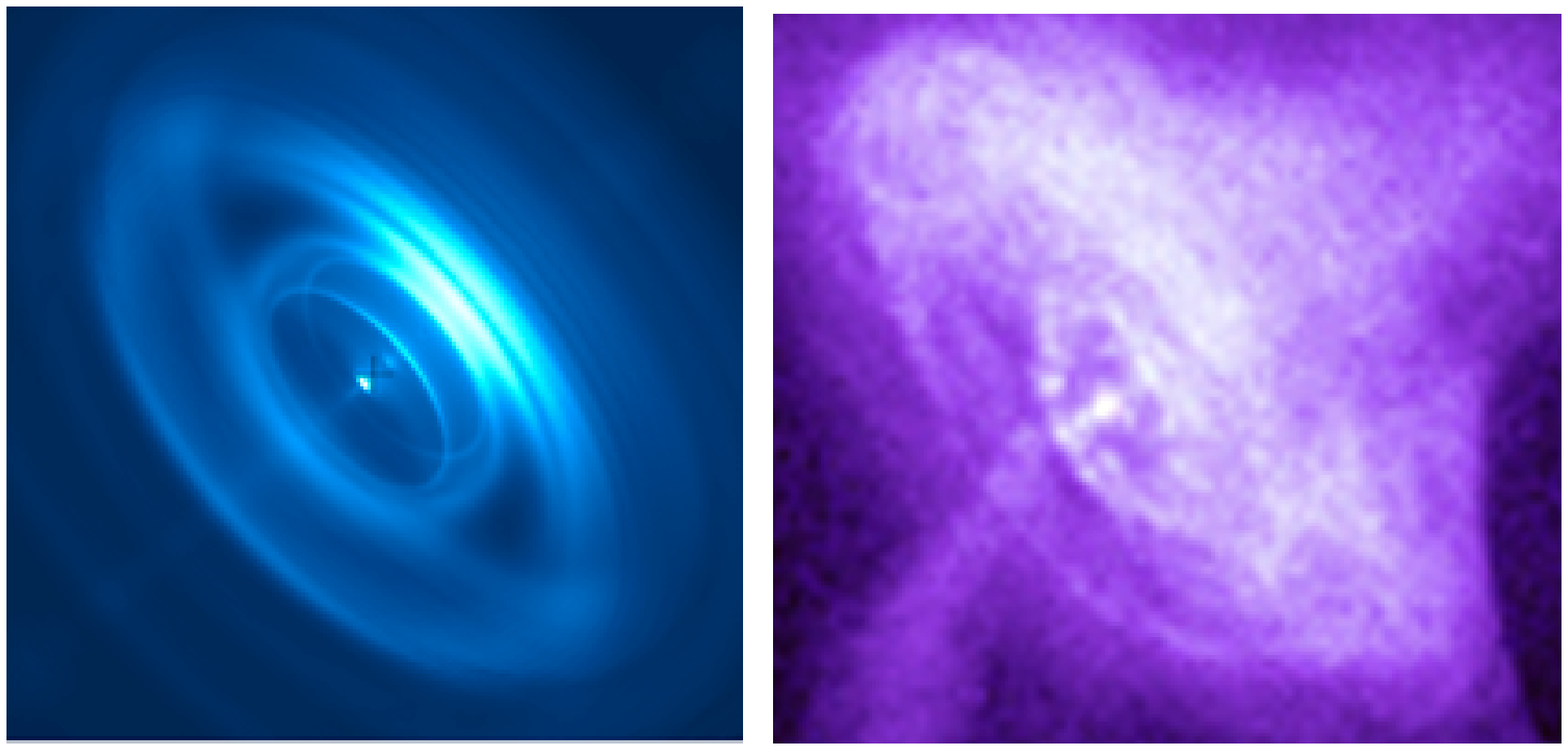}
\includegraphics[scale=.6]{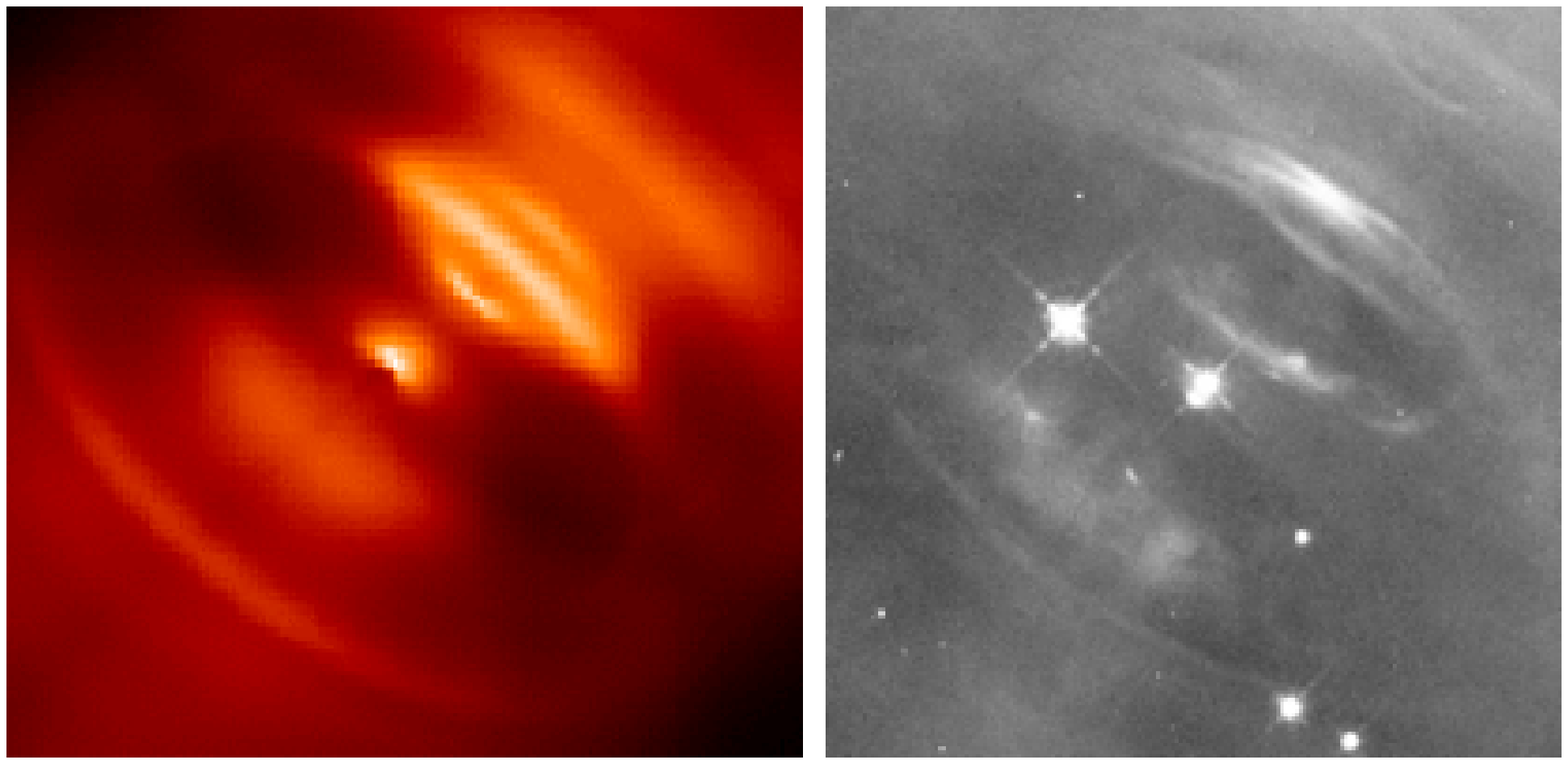}
\caption{Upper panel: left - simulated X-rays synchrotron map based on
  numerical MHD simulations of the flow \cite{cam09}; right - CHANDRA
  image of the central region of Crab Nebula. The two images have
  similar scaling. Lower panel:  left - simulated optical synchrotron map based on
  numerical MHD simulations of the flow (different from the above one); right - HST image of the wisp
region in Crab Nebula \cite{hes02}. Note the agreement between the
observed features and the results of MHD models.}
\end{figure}

As the flow expands away from the TS, toward the edge of the nebula,
the magnetization increases, until equipartition is reached. Due to
the magnetic field distribution in the wind, equipartition is first
reached close to the equator than at higher latitudes. 
The magnetic pressure prevents further compression beyond equipartition, hoop
stresses in the mildly relativistic postshock flow become efficient, 
and the flow is diverted  back toward the axis. This is the process 
that causes the formation of a collimated jet along the axis itself. 
The wind magnetization regulates the formation and properties of the
jet: for low values $\sigma < 0.001$, equipartition is not reached inside the 
nebula, and no jet is formed. At higher magnetizations  equipartition 
is reached in the close vicinity of the TS, and most of the plasma
ends in a jet. The plasma speed in the jet is $\sim 0.7 c$, in
agreement with observation of the jet in Crab Nebula and MSH 15-52
\cite{wei00,hes02,mel05,del06}, for magnetization values $\sigma\sim 0.1$. 
Associated with this collimated back-flow there is a global
circulation inside the nebula, with typical speeds $\sim 0.1 c$ that
 might lead to  mixing with cold ions \cite{lyu03}.

Numerical models offer the possibility to investigate different
distributions of
magnetic field in the wind. In particular, for oblique
rotators, while the energy distribution in the wind is identical to 
the aligned case \cite{bog01b,spi06}, the magnetic field is supposed to
give rise to a striped equatorial region, with alternating polarities.
 If this striped wind region
is dissipated, and this can happen either in the wind
\cite{lyu01b,kir03} or at the termination shock itself
\cite{lyu03b,lyu05}, then a low magnetization equatorial flow is
expected, which adds complexity to the flow structure inside the
nebula. One of the main success of the MHD model is that emission maps
based on the results of numerical simulations give different
observational signatures if an unmagnetized equatorial sector,
corresponding to the striped wind region, is present or not: a large
striped region is needed to explain the observes inner-ring
outer-torus structure of many PWNe, while models without it lead to single ring nebulae. 

When comparing observations with emission maps based on the fluid
structure derived from relativistic MHD simulations we clearly
see that, within the MHD regime, it is possible to recover almost all of
the observed features, with correct size and luminosity. However,
until now, little work has been devoted to investigate if it is possible
to discriminate among various particle injection mechanism. Work has
mostly focused on X-ray, and a uniform injection in the form of a
single power-law distribution has been
assumed \cite{ldz06,vol09,kom04,cam09}. Moreover one should be aware
that even today, with better computational facilities, it is not
possible to conduct an exaustive sampling of the entire parameter
space characterizing the interaction of PWNe with SNRs. Works has
mostly focused on reproducing Crab Nebula, where many parameters are
constrained by a rich set of observations. 

In Fig.~\ref{fig:2},  CHANDRA and HST images of the Crab Nebula are compared to 
maps based on a simulations with striped wind in X-rays and optical. The knot
and the inner ring are both present, and they are due to the high 
velocity flow in the immediate post shock region, at intermediate 
latitudes. The main torus is visible at larger distances, as well as
features like the {\it anvil} which corresponds to the backward side of the nebula.
X-ray maps of the spectral index based on simulations also agree with the
main observed properties of the Crab Nebula \cite{mor04} and with 
recent results about Vela \cite{pav07} and Kes 75 \cite{ng08}: in particular the spectrum 
appears to flatten moving away from the pulsar toward the main torus, without the need to 
assume any re-acceleration.  All this rich emission pattern is
ultimately related to Doppler boosting effects: at high speed, the emission
is enhanced (rings) and the spectrum is harder. There are still problems 
to recover the correct luminosity/spectrum in the jet. This might be
indicative of some form of dissipation and re-energization along the axis; 
possibly associated with local instabilities in the toroidal magnetic 
field \cite{beg98} which present axisymmetric simulations cannot
address, but for which there are many observational evidences 
\cite{pav03,mor04b,mel05,del06}. 

Preliminary results \cite{cam09} show that, to realistically respoduce the X-ray
emission from Crab Nebula,  $\sigma\simeq 0.1$ is required (about two
orders of magnitude higher than the 1D estimate by KC84) in
conjunction with a large $(\sim 45^\circ)$ striped zone. 

Perhaps the most promising clues to investigate the flow dynamics
inside PWNe in the future might come from polarization, and possibly 
X-ray polarization: while emission maps mostly trace the flow
velocity inside the nebula, they have little sensitivity to the 
presence of a small scale disordered component of the magnetic field.
There are several indications from optical polarimetry in the Crab
Nebula
\cite{hes08}, as well as indication from recent MHD simulations \cite{cam09,vol09},
suggesting that turbulence might be present, in the body of the
nebula, leading to a partial randomization of the field.
The effect of the flow velocity on the polarization angle has 
been discussed by \cite{me05b} and \cite{ldz06}, and the results
generally agree with available optical polarization measured in Crab
\cite{wil72,vel85,sch79,hic90,hes08}.

\subsection{Time variability}

It is known that, close
to the supposed location of the termination shock, PWNe show a short 
time variability mainly detected in optical and X-ray bands. 
This however does not affect the global properties of the 
main observed features belonging to the jet-torus structure
(i.e. the inner ring, torus and jet) which appear to
be quite persistent on long time-scales. Variability of the wisps
in the Crab Nebula has been known for a long
time\cite{hes02,bie04}. Recent observation have shown that 
the jet in Vela appears to be strongly variables \cite{pav01,pav03}, 
together with the main rings \cite{pav07}. Variability is also 
observed in the jet of Crab \cite{mor04b,mel05}, and have recently
been detected in MSH 15-52 \cite{del06}. 

In the strongly toroidal field of these nebulae, the jet variability, 
which usually has time-scale of years, is likely due to kink or
sausage mode, or even to fire-hose instability \cite{tru88}. 
On the other hand the wisps show variability on shorter time-scales of
months: the variability takes the form of an outgoing wave pattern,
with a possible year-long duty cycle.

For a long time the only model capable of reproducing the observed 
variability was the one  proposed by Spitkovsky \& Arons \cite{spi04}, based on the
assumption that ions are present in the wind. The idea of ions was
also supported by kinetic simulations of acceleration in a
strong shock \cite{ama06}.  The presence of particles with larmor
radii, of order of the size of termination shock, introduces kinetic
effects related to the separation of scales, 
leading to compression of electrons. The model however
requires a large fraction of pulsar spin-down energy in the ion
component, which contrast the basic idea of leptonic dominated
systems, and in general is not supported by spectral model of PWNe.

The most recent achievement of the MHD nebular models have been the
 ability to reproduce the observed variability
 \cite{cam09,me06b,beg99,vol09}. It is the fundamental
 multidimensional nature of the problem that allows for variability of the
 flow pattern. It was already noted in early simulations 
\cite{bog05} that the  synchrotron emissivity inside the nebula
varies. There were also evidences \cite{kom04} suggesting that the
nebular flow might have a feedback action on the TS, causing 
it to change shape, and thus inducing a change in the appearance of
the wisps.

This picture has been confirmed recently \cite{cam09,vol09}: within the MHD
regime it is possible to recover the variability, the outgoing
wave patter, its typical speed, and luminosity variations. It is found
indeed that a SASI like instability is present. Waves injected at the
termination  shock can propagate toward the axis, feeding back on the
termination shock and triggering the injection of new waves.
 Simulations show that there is a typical duty-cycle of about 1-2
 year; more generally the duty-cycle will be of the order of the
 radius of the termination shock divided by the typical propagation speed
 $\sim 0.5c$.

\subsection{Gamma rays}

At the moment the most promising observational avenue in PWNe  research is the study
of gamma-ray emission. HESS has shown that many extended gamma-ray
sources are associated with PWNe both young ones in the free expansion
 phase and older ones undergoing reverberation \cite{dej05,gal07}. New
 results from the FERMI satellite are just arriving \cite{abdo10,abdo10b},
 with data extending from the high energy MeV synchrotron part of the
 spectrum, to the 100 GeV IC part.

The emission at MeV energy observed from young object should enable us
to put constraints on the acceleration mechanism. It seems that an
exponential cutoff in energy can explain the observed data.
The more interesting question regard how the observed variability of
the wisps manifest itself at higher energies. Naively one might expect
 that the emission at high energies should show similar variability, 
on comparable time-scales, but with amplitudes of order unity.
 On the other hand preliminary results \cite{proc1}
suggest that in the 100MeV range no variability is detected. This has
important implications in term of acceleration: particles responsible
for the 100MeV emission have typical larmor radii of order of the size
of the TS. Coincidentally this is also the coherent length of the
turbulence that is at the base of the MHD variability \cite{cam09}. In this sense
high energy particles, are decoupled form the MHD flow, and their
response to the MHD turbulence is incoherent. This of course deserves
further investigation. In particular one would like to know at what
energy variability reaches a maximum, where the particles start to
decouple form the MHD flow, and to what degree this incoherent
interaction with turbulence can induce variability.  

The emission at GeV energies is assumed to be from Inverse Compton 
scattering on background radiation, and in the case of Crab the 
self synchrotron. Interestingly the spectral properties of the
 comptonized radiation can be used to derive information about 
particles which are supposed to emit synchrotron in the UV, and are
usually not directly accessible given the high UV absorption in the ISM. In the
case of MSH 15-52 \cite{abdo10b} Fermi results, have ruled out previous
EGRET measures, and shown that the undetected high energy part of the
particle distribution function must be harder than previously assumed.
Similar results could be expected for Kes 75.
GeV emission could in principle help provide an independent constraint 
on  the magnetization in the nebula.  

An alternative contribution to the gamma ray emission, if protons are
present in the pulsar wind, is the p-p scattering and related pion decay
\cite{ama03,hor06}. However uncertainties on the target number density
and the spin-down energy in protons, make this channel hard to
constrain. At present IC scattering can fit most of the objects, if
one allows for fluctuations of the local background of order unity with
respect to the galactic average. 

Recently a series of simplified evolutionary models for the high
energy emission from PWNe have been developed extending beyond
 the free expansion \cite{dejager08,gelfand09,me10}. This is quite important given that
 the majority of the gamma-ray PWNe are supposed to be
 post-reverberation objects, where the interaction with the SNR shell
 can play a major role

For example in many of these objects the bulk of the gamma emission is
 not centered on the pulsar, and the displacement is too large to be 
explained in term of a moving pulsar leaving behind a relic PWN.
 Two possible explanations have been invoked: an off-center
 compression by the reverse shock \cite{blo01}; or the formation of a
 bow-shock tail \cite{me05,kar07}, where particle responsible for
 the inverse Compton emission in the gamma-ray band can be 
advected to large distances from the pulsar. 

\section{Evolution of PWNe}
\label{sec:3}

In the previous section we have devoted our attention to the successes
of the fluid/MHD model regarding the emission properties observed in
young system. Numerical simulations offer also a way to follow the
evolution of a PWN and its interaction with the SNR at later ages. 

At the moment, however, the study of old object has been limited to a
qualitative analisys of the interaction in an attempt to recover the
main phases of the evolution and understand how the observed
multi-wavelength morphology depends on the interaction itself.

In the analytic model developed by KC84 the SNR has only a passive
role, providing the confinement of the PWN.
 Given the complexity of the PWN-SNR interaction, a detailed study of the evolution of
the system, has been possible only recently, thanks to the
improvement in computational resources \cite{van01,blo01,me03,van04,fer08}.

By comparing the energy in the SNR ($\sim
10^{51}$ ergs) to the total energy injected by the pulsar during its
lifetime ($\sim 10^{49}$ ergs) it is easy to realize that a PWN cannot significantly
affect the SNR, while the evolution of the SNR can have important
consequences for the PWN.

There are three main
phases (for a more complete discussion of PWN-SNR evolution see
\cite{ren84} and \cite{gae06}), in the PWN-SNR evolution. At the beginning the PWN expands
inside the cold SN ejecta. The SN ejecta are in free expansion, so
this phase is generally called {\it free expansion phase}.
This phase lasts for about 1000-3000 yr, and during this period the pulsar luminosity
is high and almost constant. This is the present phase of the Crab
Nebula, 3C58, MSH 15-52, G21.5, and PWNe in this phase are expected to shine in high energy
X-rays emission.  The expansion velocity of PWNe in this early stage
is typically few thousands kilometer per second. For this reason
one can neglect the pulsar kick (velocities in the range 50-300 km/s)
in modeling young objects, and assume
the pulsar to be centrally located.  As the system expands inside the
high density, cold, supersonic ejecta of the SNR, a thin shell of
swept-up material is formed. Given that the density of the shell is much higher than the
enthalpy of the relativistic plasma, the shell is subject to
Rayleigh-Taylor instability. This is supposed to be at the origin of
the filamentary network of the Crab Nebula \cite{hes96,jun98,me04b},
and 3C58 \cite{boc01}. 
In the thin-shell approximation, it has been shown that it is possible
to derive an analytic self-similar solution \cite{me04b}, describing
the expansion of the nebula in this phase.

\begin{figure}
\label{fig:5}
\begin{center}
\includegraphics[bb=0 224 305 546, clip,
  scale=.3]{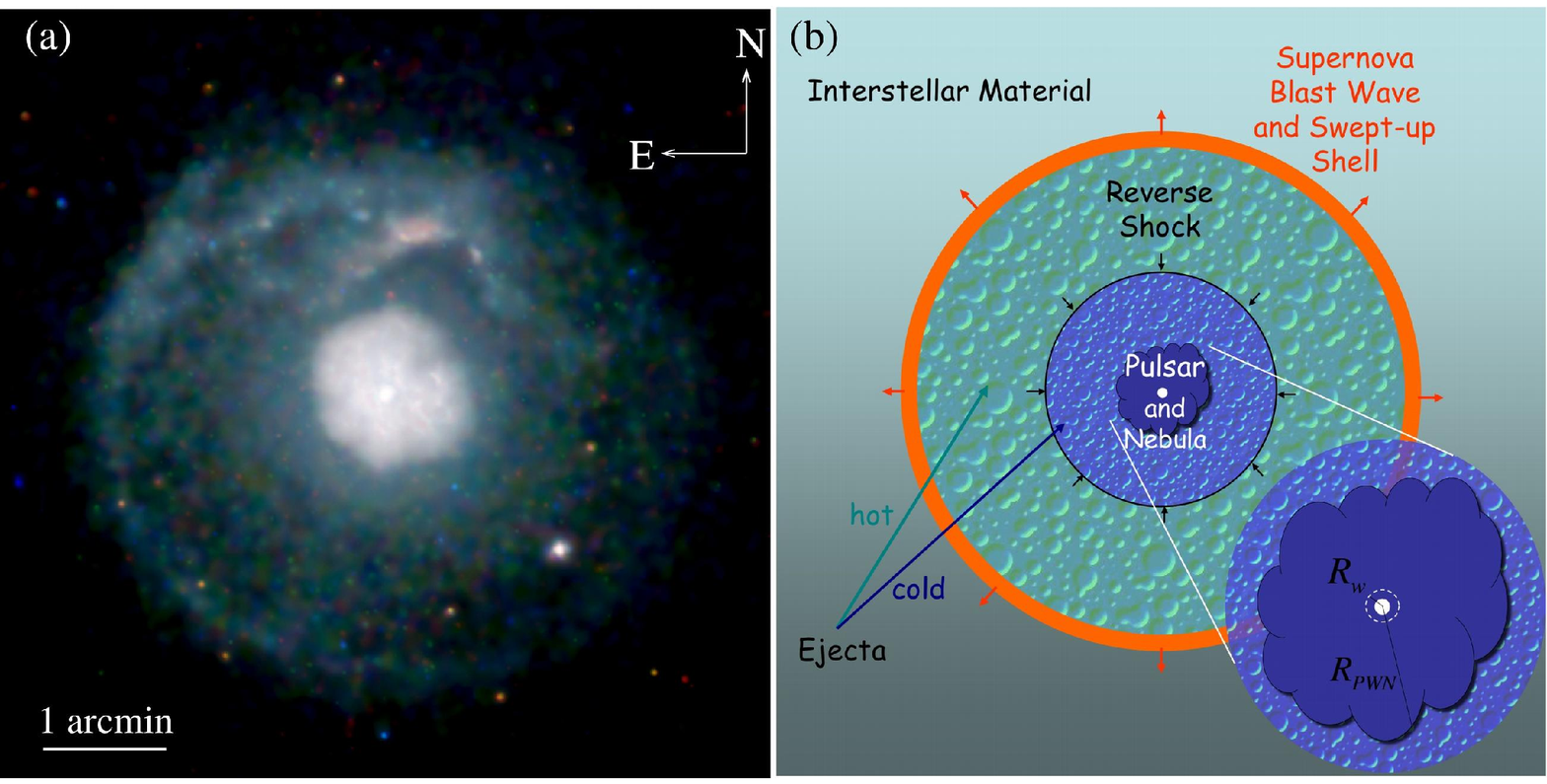}\hspace{0.5truecm}\includegraphics[bb=118 395 495
  770, angle=90,scale=0.27,clip]{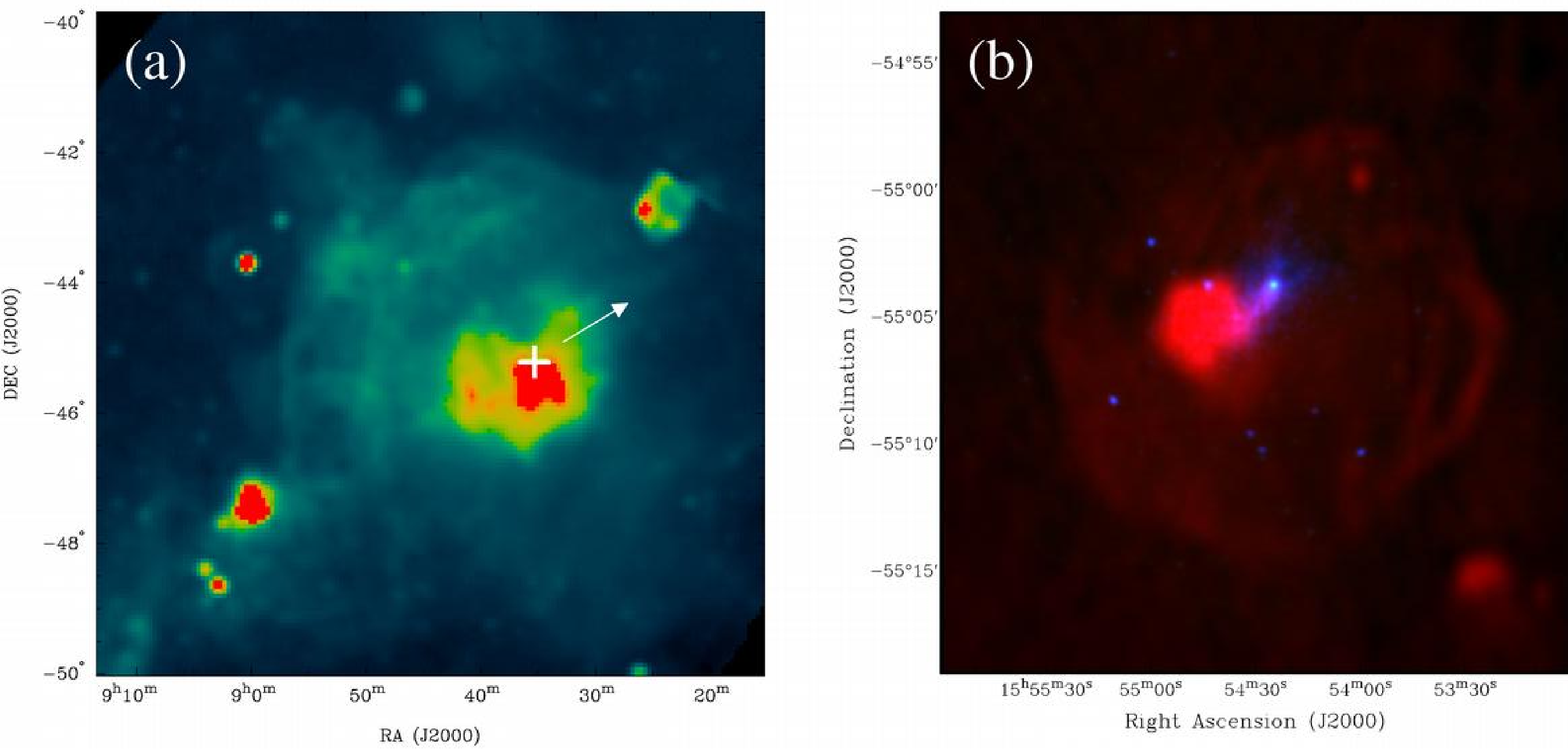}\includegraphics[bb=115 8 495
  357, angle=90,scale=0.27,clip]{gfig3.eps}
\includegraphics[scale=0.33]{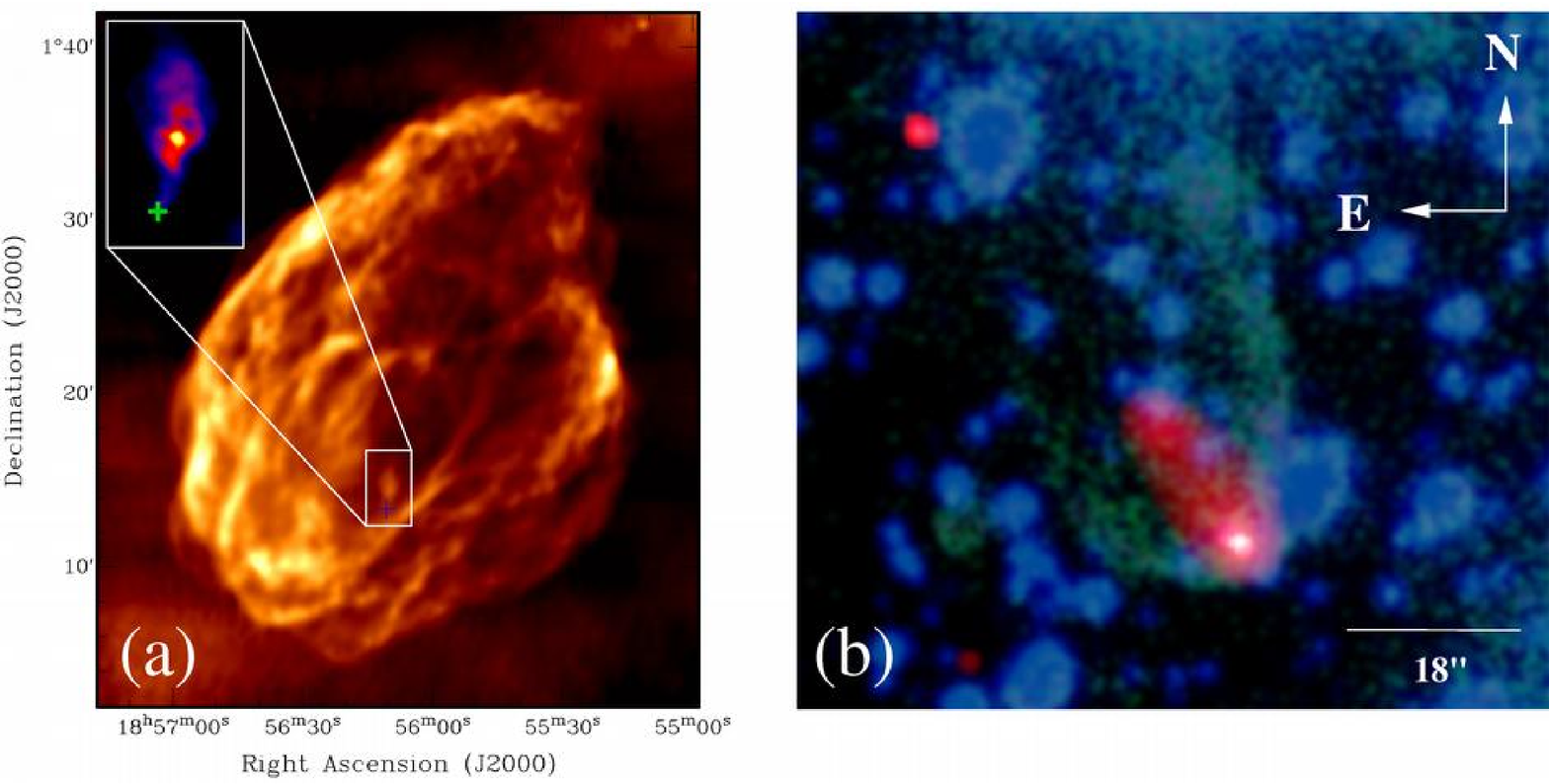}
\end{center}
\includegraphics[angle=270, scale=0.4]{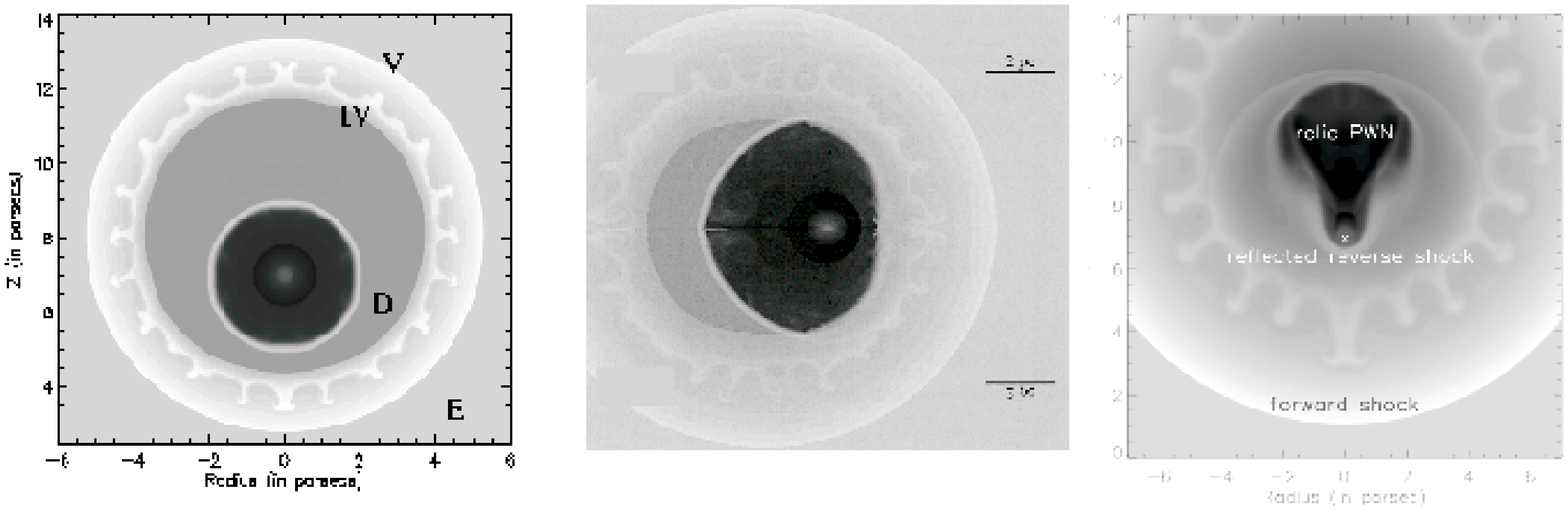}
\begin{center}
\includegraphics[bb=390 80 560 232, clip, angle=90,scale
=0.53]{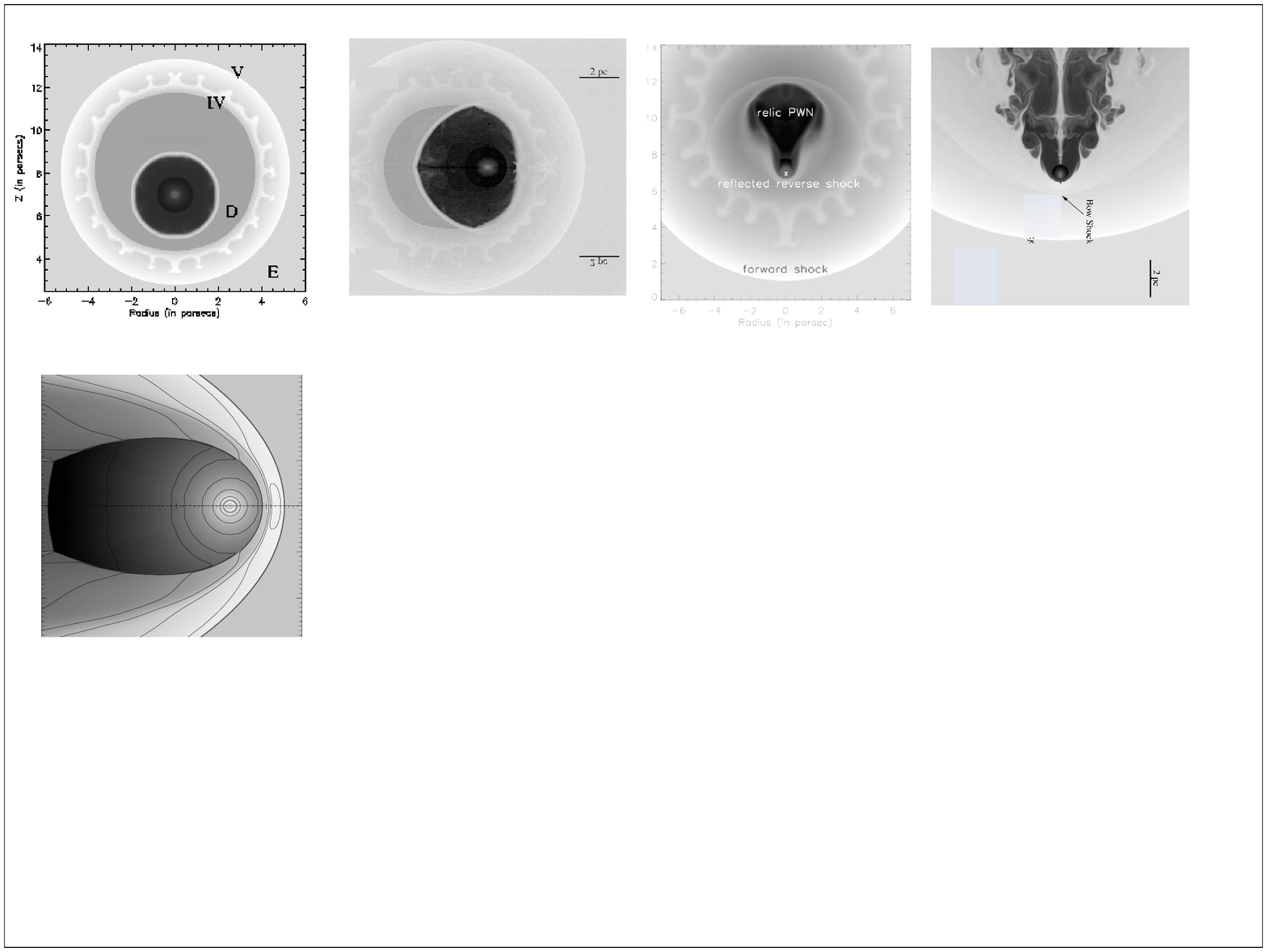}\includegraphics[bb=195 579 365 748, clip, angle=90,scale
=0.5]{model.ps}
\end{center}

\caption{Upper part: images of various evolutionary phases of PWNe
  (from Gaensler \& Slane \cite{gae06}). From Left to right, up to down: X-ray
  image of the composite remnant G21.5-0.9, free expansion phase;
  radio image of Vela SNR, displacement of the nebula due to the
  compression of the reverse shock during reverberation; SNR G
  327.1-1.1 in radio (red) and X-rays (blue), relic PWN phase; W44 in
  radio, transition to the internal bow-shock phase; PSR B1957+20 in
  H$_\alpha$ (green) an X-rays (red), ISM bow-shock nebula. Lower
  part: numerical
  hydrodynamical simulations of the various phases of the PWN-SNR
  evolution (from \cite{van04, me02}). Each figure of the
  lower part corresponds to systems shown in the upper one.}
\end{figure}

The PWN will expand until eventually it will come into contact with the reverse shock in
the SNR shell.
In the absence of a central source of energy, like a pulsar, the reverse
shock is
supposed to recede to the center of the SNR in a time of order of
5000-10000 yr \cite{tru99}. From this moment on the evolution of the PWN is
modified by the more massive and energetic SNR shell:  the PWN undergoes
a compression phase  generally referred as {\it reverberation
  phase}, that can last 5000-10000 yr. In the simple 1D scenario the
pressure in the compressed PWN will rise to balance the compression and
to push
back the ejecta, and the nebula might undergo several compression and
rarefaction cycles. This is however an artifact of the 1D geometry, and
conclusion based on the existence of these oscillations are not
reliable. More appropriate multidimensional studies have
shown that the SNR-PWN interface is highly Rayleigh-Taylor unstable
during compression \cite{blo01}, which can cause efficient mixing of
the pulsar wind material with the SNR. This mixing will most likely
prevent any oscillation and the system might rapidly relax to pressure
equilibrium. This reverberation phase is
supposed to last about $10^4$ years. Even if energy injection
from the pulsar at these later times is negligible, PWNe can still be
observed, due to the re-energization during compression. The
interaction with the reverse shock can lead to a variety of different
morphological structures if one consider also the pulsar proper motion
\cite{van03b,van04,me05,fer08}. The most likely outcome of the interaction
is that the nebula can be displaced with respect to the
location of the pulsar. At the beginning this might result into a
system where the pulsar is not located at the center of the radio non
thermal emission, analogous to what is observed in Vela. As the system
evolves the reverse shock will completely displace the body of the
PWN, creating a relic nebula. The relic PWN will mostly contain low
energy particles, and will be visible in radio, while high energy
particles, observable in X-rays will be seen only close to the
pulsar. G327.1 shows indeed this kind of morphology \cite{gae06}. 
Depending on projection effects one might also end up with very small
X-ray nebulae centered on vast and large radio nebulae. In this regard
polarization might prove essential to disentangle the structure. In
particular, if the Rayleigh-Taylor instability is efficient and the
radio nebula is disrupted and mixed with the ejecta, one expects to
find a low level of polarization and no evidence for a global toroidal
field. To this one must
add the possibility that the ISM magnetic field might be dragged inside
the PWN, or the nebular field inside the SNR shell. Given the importance
of magnetic configurations for particle (cosmic ray) diffusion, one
understands how important a proper model of older objects is.

At later time the SNR ejecta starts cooling and the pulsar will eventually
become supersonic. Once this happen the pulsar will form around itself a bow-shock
PWN, and one expects an emission tail to form connecting the pulsar to
the relic PWN. This model apply to the morphology and structure of W44
\cite{gae06}, or possible IC443. Interestingly, the
location with respect to the SNR where this happens does not depend on
the pulsar proper motion, and turns out to be $\sim 70$\% of the radius
of the forward shock. An obvious question is if the jet-torus
structure, that is observed in young and relatively undisturbed systems,
can survive the later interaction with the SNR shell and the reverse
shock. We know that in Vela the jet-torus is visible and does not
appear to be distorted, this suggest that, as long as the evolution is
subsonic, in the pulsar vicinity the dynamic of the nebula flow will
still be regulated by the pulsar wind. On the other hand bow-shock
simulations have shown that the typical size of the nebula in the head
of the bow-shock and the flow dynamics do not allow the formation of
collimated structures \cite{me01,me05,vig07}. In the case of  SNR G327.1-1.1 
\cite{tem09} have suggested the presence of a jet-torus, but
photons count and resolution are not high enough to make a definite
statement. Deeper observation of transitional objects are needed.

The ultimate phase of a PWN evolution depends on the pulsar kick. For
slow moving pulsars the PWN will expands adiabatically inside the
heated SNR, now in Sedov phase. Given the absence of  energy
injection, a PWN in this stage is probably only observable as a faint
extended radio source, or possibly as a large TeV nebula due to IC
from the relic leptons. To some extent this relic particle population
might contribute to the diffuse gamma-ray background.
On the contrary a fast moving pulsar can escape from the
SNR, and will give rise to a bow-shock nebula due to the interaction with the
ISM, through which it is moving at supersonic speeds
\cite{me01,me05,vig07}. These objects are observed both in $H_\alpha$
emission, due to ionization of ISM neutral hydrogen, and as long
extended cometary-like source of non thermal radio and X-ray emission, due to the
shocked pulsar wind, now forced to flow in the direction opposite to
the pulsar motion. These nebulae might constitute one of the primary
sources of positrons in the galaxy.

Bow-shock PWNe, constitute a very interesting class among PWNe, and
have recently received some attention, in particular regarding their
X-ray emission. MHD models predict that the outflow in the tail of the
bow-shock should have high speeds, of order of $0.5 c$, and that the
tail should form a very well collimated channel, with cross section
comparable with the bow-shock size \cite{me01,me05,vig07}. However observations have shown
that in general the tail is wider than expected and that typical flow
speed are high but of order $10000 $ km/s. It has been suggested that
some form of mixing with the ISM, either via shear instability between
the fast relativistic tail and the surrounding slower ISM, or via some
particle contamination by ionized neutrals coming from the ISM, might
be at play. A more detailed study of the fate of pairs injected in
the tails of bow-shock nebulae, is essential, to assess the
importance of pulsar as contributor to the pair CR background.

\section{Conclusion}
\label{sec:concl}

In the last few years, the combination of high resolution
observations, and numerical simulations, has improved our
understanding of the evolution and internal dynamics of PWNe. We can
reproduce the observed jet-torus structure and we can relate  the
formation of the jet in the post shock flow to the wind
magnetization. Simulated maps can reproduce many of the observed
features, including the details of spectral properties. Results
suggest that, the best agreement is achieved in the case of a 
wind with a large striped zone, even if MHD simulations are not able to distinguish between
dissipation of the current sheet in the wind or at the TS. Results
also suggest that it is  possible to use X-rays imaging to constrain
the pulsar wind properties; already the rings and tori observed in
many PWNe have been used to determine the spin axis of the pulsar
\cite{rom05}. Interestingly in Crab the inner ring appear less 
boosted in X-rays than the optical wisps (which shoud trace the same
flow structure) are.

Despite the undeniable successes of the MHD model, which are
universally recognized within the community, and the fact that there
is general agreement that the observed X-ray properties are strongly
dependent on the internal dynamic at the termination shock, for reasons
unknown to the author, the old KC84 model is still used as a canonical
reference for interpreting observations. While this might be
understandable for Radio or Optical data (where emission is quite
homogeneous), it is completely unreasonable in X-ray. The fact that
KC84 model is analytic and simple, is no excuse for its use, when it
is clear that it is both qualitatively and quantitatively wrong, to
the point that it basically fails to explain almost every single
aspect of X-ray data. Moreover numerical tools and facilities are today
widely available to conduct a correct study. 

Interestingly, even when data are unreasonably averaged over spherical shells (even
if imaging shows no hint of sphericity), they cannot be modeled using
the original KC84 structure, and arbitrary velocity profiles are often
assumed to reproduce the data. So one trades a model (KC84) which
is wrong (nebulae are far from spherical), but at least dynamically
consistent (the correct solution
of MHD equations), for  models which not
only are wrong but also dynamically inconsistent, to the point that
informations derived in this way have zero scientific valence, and the
entire analisys is nothing more than a fit to the data.

There are still however unsolved questions, and possible future
developments for research in this field.  All present simulations are
axisymmetric, and none is able to address the problem of the stability
of the toroidal field, nor can they reproduce the observed emission
from the jet.  It is not clear if small scale disordered field is
present in the inner region: maybe a residual of the dissipation in
the TS of the striped wind, or an outcome of the turbulence injected
by the SASI-like instability of the TS. A combination of simulations and
polarimetry might help to answer this question.

Perhaps the more interesting developments might come from either the
study of old systems, or from the investigation of particle energy
distribution signatures in young ones. For the former, what is really
needed is a large parameter study, where the various effects of pulsar
proper motion and SNR reverberation are taken into account together
with some simplified treatment of the spectral properties of these
nebulae, in order to  go beyond the simple qualitative morphological
agreement, and provide templates for emission and spectral properties to
be compared with observations. For the latter, a study of possible
signature of different injection mechanism at the TS, either as a
function of shock properties, latitude or time, should be carried on
by following the full particle distribution function in the nebula,
instead of the simple power-law assumption. This will allow to verify
if observable signature should be expected, in what band, and provide
some constrain on the physics at the TS, and more important, on
relativistic shock acceleration in general.

\begin{acknowledgement}
N.B. was supported by a NORDITA Fellowship grant.
\end{acknowledgement}
%

%
%
%

\end{document}